\newcommand{\M}{{\cal M}}
\newcommand{\s}{{\cal S}}

\documentclass[draft, numberedheadings]{aipproc}
\layoutstyle{6x9}

\title[Symmetric Criticality]{Symmetric Criticality in Classical Field Theory\thanks{To appear in the proceedings of the XIX International Fall Workshop on Geometry and Physics.}
} 
\author{C.~G.~Torre}{address={Department of Physics, Utah State University, 84322-4415, USA}}

\begin{document}
\date{November 2010}

\begin{abstract}
This is a brief overview of work done by Ian Anderson, Mark Fels, and myself on  symmetry reduction of Lagrangians and Euler-Lagrange equations, a subject closely related to Palais' Principle of Symmetric Criticality.   After providing a little history, I describe necessary and sufficient conditions on a group action such that reduction of a group-invariant Lagrangian by the symmetry group yields the correct symmetry-reduced Euler-Lagrange equations.  
\end{abstract}

\maketitle

\section{Introduction}

The following is an informal overview of what is known about symmetry reduction of Lagrangians and Euler-Lagrange equations, a subject closely related to Palais' Principle of Symmetric Criticality \citep{Palais:1979}.  To give you a feel for what I have in mind, let me begin by reviewing a very simple, probably quite familiar example.

Suppose you are searching for spherically symmetric solutions of the Laplace equation in a spherical region $M\subset {\bf R}^{3}$ centered at the origin. These are functions $\varphi(x,y,z)$ satisfying
\begin{equation}
\label{Lap}
\frac{\partial^{2}\varphi}{\partial x^{2}} + \frac{\partial^{2}\varphi}{\partial y^{2}} + \frac{\partial^{2}\varphi}{\partial z^{2}} = 0,
\end{equation}
and which are invariant under the standard action of $SO(3)$ on ${\bf R}^{3}$.  An obvious way to proceed is to note that all spherically symmetric functions depend upon $(x,y,z)$ only through a function $q$ of the radius $r = \sqrt{x^{2} + y^{2} + z^{2}}$:
\begin{equation}
\label{inv}
\varphi(x,y,z) = q(r).
\end{equation}
A simple computation then shows that the Laplace equation restricted to spherically symmetric functions becomes
\begin{equation}
\label{redLap}
q^{\prime\prime} + \frac{2}{r} q^{\prime} = 0,
\end{equation}
which is easily solved.  Thus every rotationally invariant solution of the Laplace equation is determined by a function $q$ of one variable satisfying the ODE \eqref{redLap}.

This is, of course, an elementary example of using symmetry reduction to find group invariant solutions of differential equations \citep{Olver:1993, AFT:2000}.  Briefly, we have a group $G=SO(3)$ acting on $M$ and on the functions $\varphi$ on $M$, and we have a differential equation \eqref{Lap} which is invariant under $G$.  The $G$-invariant solutions to \eqref{Lap} are determined via \eqref{inv} by a function $q$ on $M/G \subset {\bf R}^{+}$  -- the {\it reduced field} --  which satisfies a differential equation \eqref{redLap} on $M/G$, which we call the {\it reduced equation}.

An alternate approach to finding the rotationally invariant solutions to \eqref{Lap} is to observe that the Laplace equation is equivalent to the Euler-Lagrange equation associated with the critical points of the action integral:
\begin{equation}
\label{S}
A[\varphi] = \frac{1}{2}\int_{M} dV\, ||\nabla\varphi||^{2}.
\end{equation}
Rather than evaluating the field equation \eqref{Lap} on the $G$-invariant field \eqref{inv}, consider evaluating the action integral \eqref{S}  on the $G$-invariant field $\varphi(q)$ defined in \eqref{inv}. This leads to a {\it reduced action} $\hat A$ and {\it reduced Lagrangian} $\hat L$ for the reduced field $q$ on $M/G$:

\begin{equation}
\label{Shat}
\hat A[q] \equiv A[\varphi(q)]  = 2\pi \int_{M/G} dr\, r^{2}\, q^{\prime 2} = \int_{M/G} \hat L.
\end{equation}
The Euler-Lagrange equation for $q$  defined by $\hat L$ is equivalent to \eqref{redLap}, so the critical points of the symmetry reduced action also give the rotationally invariant solutions to \eqref{Lap}. 

Evidently, in this example we can obtain the reduced equation by either of two methods.  We can compute the Euler-Lagrange equation \eqref{Lap} of the action \eqref{S} and symmetry-reduce the result, {\it or} we can  symmetry reduce the action \eqref{S} and compute the Euler-Lagrange equation of the result \eqref{Shat}.  Both methods lead to the same reduced equation \eqref{redLap}.  

The foregoing discussion provides a simple illustration of the idea that ``symmetric critical points are critical symmetric points'', which is Palais' Principle of Symmetric Criticality \citep{Palais:1979}. When the principle is valid, one may perform symmetry reductions at the level of the action, not just at the level of the field equations. There are a number of reasons why one would desire a symmetry-reduced variational principle.  Firstly, there is the pragmatic consideration of computational expediency: it is usually simpler to symmetry reduce a Lagrangian and compute the resulting reduced equations than it is to symmetry reduce the field equations directly.  But, perhaps more importantly, in physics one is frequently using symmetry reductions of field theories to extract simplified models which are amenable to analysis. The existence of a symmetry-reduced variational principle for such models is advantageous since it defines a link between symmetries and conservation laws for the model, it defines symplectic and Hamiltonian structures for the model, and can be used to formulate the quantum theory of the model.  For example, see the recent review article on quantization of symmetry-reduced versions of general relativity \citep{lrr-2010-6}.

As Palais emphasizes, the Principle of Symmetric Criticality need not be well defined, and even when it is well-defined it is {\it not} always valid.  In particular, there is no {\it a priori} reason why the symmetry reduction of a Lagrangian must yield Euler-Lagrange equations equivalent to those governing $G$-invariant solutions of the original Euler-Lagrange equations.   
My purpose here is to review the problem of symmetry reduction of fields, field equations, and Lagrangians, and to describe the results of Ian Anderson, Mark Fels, and myself \citep{AFT:2000, AF:1997, Anderson:1999cm, FT:2002}, which provide necessary and sufficient conditions on a symmetry group such that the symmetry reduction of any Lagrangian  yields the (correct) symmetry-reduced field equations.

\section{Some History}

The history of symmetry reduction of variational principles and the principle of symmetric criticality is a long one, principally concentrated in the literature on general relativity and gravitation.  In 1917 Weyl re-derived the Schwarzschild solution to the vacuum Einstein equations by imposing spherical symmetry on the metric in the Einstein-Hilbert action and solving the resulting Euler-Lagrange equations \citep{Weyl:1917}.\footnote{Strictly speaking, Weyl went beyond spherical symmetry in that he did not use the most general spherically symmetric metric, having restricted the form of the metric with some additional coordinate conditions. Thus it is even more remarkable that this shortcut to the reduced equations worked at all!}  Half a century later, in 1972, Lovelock took note of this remarkable state of affairs and pointed out that a number of diffeomorphism invariant Lagrangians  would allow for the same successful symmetry reduction of the variational principle \citep{Lovelock:1973}.  We shall see below that reductions by spherical symmetry will {\it always} work, irrespective of the choice of Lagrangian.

In 1969 Hawking found a class of symmetry reductions where one cannot successfully reduce the Einstein-Hilbert variational principle \citep{Hawking:1969}.  He was considering spatially homogeneous cosmological models, obtained by fixing a three-dimensional connected Lie group $G$ and considering spacetimes of the form $R\times G$, where the isometry group is $G$ acting on itself, say, from the left.  Using the Maurer-Cartan forms on $G$, Hawking constructed the most general $G$-invariant metric, which is characterized by 10 freely specifiable functions of one variable (time). He substituted this general $G$-invariant metric into the Einstein-Hilbert action and obtained a symmetry-reduced variational principle for the 10 unknown functions. He found that the equations of motion coming from this reduced variational principle coincided with the Einstein equations only if the isometry group $G$ had a Lie algebra of type A in the Bianchi classification.  This result  (which, it should be pointed out, was not the main point of Hawking's paper) was amplified and studied in more detail by MacCallum and Taub \citep{MacCallum:1972}, Ryan \citep{Ryan:1974}, Sneddon \citep{Sneddon:1976}, and others.  

So, it became clear that symmetry reduction of variational principles is somehow unreliable, although at this stage it was not quite clear how best to characterize the problem.  Is the problem with the Lagrangian? Is it with the choice of symmetry group?   What are necessary and sufficient conditions for a successful symmetry reduction of a variational principle? In much of the literature described above the emphasis was placed on the success or failure of symmetry reduction of a particular Lagrangian ({\it e.g.}, the Einstein-Hilbert Lagrangian), which tended to cloud the issue.  The contribution of Palais in 1979, his ``principle of symmetric criticality'',  provided a more profitable viewpoint in which attention is focused on specific properties of the symmetry group as determining whether or not one can reduce a generic variational principle. 

\section{Palais' Principle of Symmetric Criticality}

Consider a manifold $\M$ upon which a group $G$ acts.  Let $\s\subset \M$ be the points which are fixed by $G$, that is, $\s = \{x\in\M | g\cdot x = x, \forall g\in G\}$. We assume $\s$ is a submanifold with embedding $i\colon\s\to\M$. Let $A\colon \M \to {\bf R}$ be any $G$-invariant function on $\M$, that is, $A\circ g = A$, $\forall g\in G$.  Palais' Principle of Symmetric Criticality (PSC) asserts that a necessary and, more importantly, a sufficient condition for $p\in \s$ to be a critical point of $A$ is that the derivatives of $A$ in directions tangent to $\s$ all vanish:

\begin{equation}
d(i^{*}A)\Big|_{p} = 0 \quad\Longleftrightarrow\quad dA\Big|_{p} = 0.
\end{equation}

The relevance of PSC to the symmetry reduction of variational principles in field theory becomes clear if one makes the following identifications: $\M$ corresponds to the space of fields, $G$ corresponds to the symmetry reduction group,  $\s$ corresponds to the set of group-invariant fields, $A$ corresponds to the action, and $i^{*}A$ corresponds to the reduced action.

Palais was quick to point out that PSC need not be well-defined, {\it e.g.}, if $\s$ is not a manifold, and even when it is defined it may not be valid, as Hawking found with spatially homogeneous cosmological models.  However, in situations where PSC makes sense, Palais provides some necessary and sufficient conditions for PSC in the setting where  $\M$ is a Banach $G$-manifold modeled on a Banach space $\cal V$. Two important sufficient conditions for PSC to be valid include (i) $\M$ is a Riemannian manifold and $G$ acts isometrically on $\M$, and (ii) $G$ is compact.  A necessary and sufficient condition for PSC arises if one makes a further assumption that the action of $G$ on $\M$ is {\it linearizable}. This means that there exists a coordinate chart in the neighborhood of any point $p\in\M$ such that the action of $G$ is the restriction of a linear transformation on ${\cal V}$  to the open set used to define the chart.  One can now use the action of $G$ on ${\cal V}$ to characterize the validity of PSC.  Let $\Sigma\subset {\cal V}$ be the set of $G$ invariant vectors.  By duality, we have a (linear) action of $G$ on ${\cal V}^{*}$; let $\Sigma_{*}\subset {\cal V}^{*}$ denote the $G$ invariant dual vectors. Finally, let $\Sigma^{0}\subset {\cal V}^{*}$ denote the annihilator of $\Sigma$, that is, the set of linear functions on ${\cal V}$ which map all elements of $\Sigma$ to zero.  I will define the {\it Palais condition} to be 
\begin{equation}
\label{Palais}
\Sigma_{*} \cap \Sigma^{0} = 0.
\end{equation}
For a linearizable $G$ action on a Banach $G$-manifold $\M$, the Palais condition  is necessary and sufficient for PSC to be valid for all $G$ invariant functions on $\M$.

One important feature to notice here is that the validity of PSC has been reduced to a question about the action of the symmetry group on $\M$.  This comes about because one has demanded that PSC be valid for all $G$-invariant functions.  If the Palais condition is satisfied, then one is guaranteed that symmetric critical points are always critical symmetric points irrespective of the choice of the ($G$-invariant) function $A$.

It is possible to give a more detailed and in many ways simpler condition for PSC, provided one restricts the setting to local Lagrangian field theories.  This is the subject of the next section.

\section{A Local Version of PSC}

The following represents work by Ian Anderson, Mark Fels, and myself.  One outcome of this work is a reformulation of PSC within the context of local Lagrangian field theory.  This setting is mathematically simpler than the very general setting (Banach manifolds) in which Palais performed his investigation. Consequently, one can give a simple,  detailed and easily verifiable set of necessary and sufficient conditions for the validity of PSC. A key feature of this local version of PSC is that attention is focused away from critical points {\it per se} in favor of the differential equations defining the critical points. This allows one to avoid consideration of function spaces, boundary conditions, the structure of the set $\cal S$ of symmetric points, existence of  critical points, {\it etc.}  Consequently, the version of PSC discussed in the following should not, strictly speaking, be considered a specialization of Palais' PSC, although there will be situations where they agree.

The local version of PSC involves symmetry reduction at 3 levels: reduction of fields, reduction of differential equations, and reduction of Lagrangians. We summarize the key points from each in what follows.  For details, see \citep{AF:1997, AFT:2000}.

\subsection{Reduction of Fields}

Fields will be taken to be smooth cross sections $\varphi\colon M\to E$ of a fiber bundle $\pi\colon E\to M$ upon which there is a smooth, projectable group action $\mu\colon G\times E \to E$. The group action,  by projection,  induces a transformation group on the $n$-dimensional manifold $M$; we assume all the group orbits in $M$ have dimension $l$. The group acts by pullback  on the set of fields; we denote the action of $g\in G$ upon $\varphi$ by $\mu_{g}\varphi$.  For each $x\in M$  denote by $G_{x}\subset G$ the isotropy group of $x$. $G_{x}$ is the subgroup of $G$ which fixes $x$.  For the fields typically found in physics, the group will usually not act transversely to the fibers of $E$ if the isotropy groups are non-trivial.\footnote{I should point out that much of the mathematical literature on the theory of symmetry reduction of differential equations assumes the group action on $E$ is transverse.}

We now consider $G$-invariant fields, which are the fields satisfying $\mu_{g}\varphi = \varphi$, $\forall g\in G$.  A key observation is that the $G$-invariant fields are (for sufficiently well-behaved group actions) cross sections of a sub-bundle $\pi\colon \kappa(E) \to M$ upon which  the group acts transversely. The fiber of $\kappa(E)$ at each $x\in M$ is determined by invariance of the fields with respect to the isotropy $G_{x}$.  One can take the quotient of $\kappa(E)$ by the group action to obtain a {\it reduced bundle}, $\hat \pi \colon \kappa(E)/G \to M/G$.  We make the blanket assumption that the reduced bundle is a smooth fiber bundle. It can be shown that there is a one to one correspondence between sections $q\colon M/G \to \kappa(E)/G$ and $G$-invariant sections of $E$. We henceforth denote the $G$-invariant sections of $E$ by $\varphi(q)$.  To summarize, the analysis of $G$-invariant fields can always be formulated in terms of {\it reduced fields}, which are  defined as sections $q\colon M/G \to \kappa(E)/G$.

\subsection{Reduction of Field Equations}

We will restrict attention to field equations arising as Euler-Lagrange equations and freely use the terminology and notation of the variational calculus. If $M$ is $n$-dimensional, the Lagrangian will be viewed as an $n$-form $\lambda[\varphi]$ locally constructed from the field $\varphi$ and its derivatives to some finite order.\footnote{Put differently, we view the Lagrangian as a mapping from the jet space of sections of $E$ into the space of $n$-forms on $M$.} We have the fundamental formula for the variation of the Lagrangian
\begin{equation}
\label{deltalambda}
\delta \lambda[\varphi] = E(\lambda)[\varphi] \cdot \delta \varphi + d\eta[\varphi, \delta\varphi].
\end{equation}
The field equations are $E(\lambda)[\varphi] = 0$ and the $(n-1)$-form $\eta$ will be referred to as the {\it boundary form}. Note that $\eta$ is only determined by $\delta\lambda$ up to the addition of an exact form locally constructed $\varphi$ and $\delta\varphi$ (and their derivatives), the latter occurring linearly.

We assume $\lambda$ is invariant under the action of $G$ on $\varphi$ in the sense that
\begin{equation}
\label{laginv}
\lambda[\mu_{g}\varphi] = \mu_{g}^{*} \lambda[\varphi], \quad \forall g\in G.
\end{equation}
This implies the field equations are $G$-invariant as well:
\begin{equation}
\label{eqinv}
E(\lambda)[\mu_{g}\varphi] = \mu_{g}^{*} E(\lambda)[\varphi],\quad \forall g\in G.
\end{equation}

Now we consider symmetry reduction, where we restrict the field equations $E(\lambda)[\varphi] = 0$ to $G$-invariant fields $\varphi(q)$.  The field equations for $G$-invariant fields can be expressed as  a system of differential equations $\Delta[q]=0$ for the reduced fields $q$ \citep{Olver:1993, AFT:2000}.  These are the {\it reduced equations}, which  are equivalent to $E(\lambda)[\varphi(q)] = 0$.
Thus $G$-invariant field equations for $G$-invariant sections of $\pi\colon E\to M$ descend to define differential equations for sections of $\hat\pi\colon \kappa(E)/G \to M/G$.  The issue at hand is whether the Lagrangian $\lambda$ will descend to a Lagrangian for the fields $q$ on $M/G$ with Euler-Lagrange equations which are equivalent to the reduced equations $\Delta[q]=0$.

\subsection{Reduction of Lagrangians}

Consider evaluating  a $G$-invariant Lagrangian $\lambda[\varphi]$ on the set of $G$-invariant fields $\varphi(q)$.   We expect $\lambda[\varphi(q)]$  to determine a Lagrangian for the reduced fields $q$ on $M/G$.   The only slight subtlety here is that $M/G$ is an $(n-l)$-dimensional manifold, where $l$ is the dimension of the orbits of $G$ in $M$. The Lagrangian $\lambda$ for fields $\varphi$ on $M$  is an $n$-form, while the Lagrangian for fields on $M/G$ should be an $(n-l)$-form --- how to drop $\lambda$ from $M$ to $M/G$?  We do this as follows \citep{AF:1997, FT:2002}. 

Let $\chi$ be a skew-symmetric tensor of type $\left(l\atop0\right)$ which  is everywhere tangent to the orbits of $G$ in $M$. (Such a tensor field can be constructed from tensor products of the vector fields which generate the action of $G$ on $M$.)  We suppose that  $\chi$ can be chosen to be $G$-invariant. We define 
\begin{equation}
\tilde\lambda[q] = \chi\cdot \lambda[\varphi(q)],
\end{equation}
where the dot indicates contraction of $\chi$ with the first $l$ arguments of the $n$-form $\lambda$.  From \eqref{laginv} it follows that $\lambda[\varphi(q)]$ is a $G$-invariant $n$-form on $M$. Consequently, the $(n-l)$ form $\tilde\lambda$ is also $G$-invariant. Moreover, it satisfies $\xi \cdot \tilde\lambda = 0$, where $\xi$ is any vector field on $M$ generating a one parameter subgroup of the action of $G$ on $M$.  Therefore there exists a unique $(n-l)$-form $\hat\lambda[q]$ on $M/G$ defined by 
\begin{equation}
\tilde\pi^{*}\hat\lambda[q] = \tilde\lambda[q],
\end{equation}
where $\tilde\pi\colon M\to M/G$ is the projection onto the set of orbits of $G$.  $\hat\lambda[q]$ is the {\it reduced Lagrangian} for the fields $q$ on $M/G$.   

This procedure which used the tensor field $\chi$ to reduce the $G$-invariant $n$-form $\lambda[\varphi(q)]$ on $M$ to the $(n-l)$-form $\hat\lambda[q]$ on $M/G$ we call the {\it reduction map} and we denote it by $\rho_{\chi}$,
\begin{equation}
\hat\lambda[q] = \rho_{\chi}\left(\lambda[\varphi(q)]\right).
\end{equation}
It is not hard to see that the reduction map can be extended to any $G$-invariant differential form on $M$ locally constructed from $\varphi$ and its derivatives.

\section{Local Version of PSC}

We are now ready to formulate a field theoretic version of the Principle of Symmetric Criticality. For a given $G$ action on $E$, the principle asserts that around each $x\in M$ there is a $G$-invariant neighborhood upon which is defined a $G$-invariant $\chi$ such that, for {\it any} $G$-invariant Lagrangian $\lambda[\varphi]$, the Euler-Lagrange equations for $q$ coming from the reduced Lagrangian, $E(\hat\lambda)[q]=0$,  are equivalent to the reduced equations $E(\lambda)[\varphi(q)]=0$.

I must emphasize that PSC is a property of a symmetry group and is not a property of a specific Lagrangian. It is possible to have a particular Lagrangian that yields a correct reduced Lagrangian for some symmetry reduction even if PSC fails in the above sense.  As an extreme example, the Lagrangian $\lambda=0$ will always yield a correct reduced Lagrangian even if PSC is not valid for the given group action. The point of PSC is that it guarantees the reduced equations coming from {\it any} $G$-invariant Lagrangian will be equivalent to the Euler-Lagrange equations of the corresponding reduced Lagrangian.\footnote{Note that one can usually conjure up Lagrangians for which the Euler-Lagrange equations have no solution! One nice feature of the local version of PSC being discussed here is that, by focusing attention on equivalence of differential equations, one avoids having to worry about whether the parent variational principle has any critical points --- not to mention that such existence properties may be inconvenient to establish.}

There are two obstructions to the validity of PSC in this setting.  They can be understood by noting that the variational formula for $\delta\hat\lambda[q]$, 
\begin{equation}
\label{deltahatlambda}
\delta\hat\lambda[q] = E(\hat\lambda)[q]\cdot \delta q + d\hat\eta[q, \delta q],
\end{equation}
can be computed by applying the reduction map to  $\delta\lambda[\varphi]$:
\begin{equation}
\delta\hat\lambda[q] = \rho_{\chi}\left(\delta\lambda[\varphi(q)]\right).
\end{equation}
We then have
\begin{equation}
\label{var}
\tilde\pi^{*}\left(E(\hat\lambda)[q]\cdot \delta q + d\hat \eta[q, \delta q]\right) = \chi\cdot \Big(E(\lambda)[\varphi(q)]\cdot \delta \varphi(q)\Big) +  \chi\cdot d\eta[\varphi(q), \delta\varphi(q)].
\end{equation}
The Euler-Lagrange expression $E(\hat\lambda)[q]\cdot \delta q$ is determined by all terms on the right hand side of \eqref{var} which are not contained in an exact form. The boundary term $d\hat\eta$ is determined by all the terms on the right hand side of \eqref{var} which appear via an exact form. From \eqref{var} it follows that in order for $E(\hat\lambda)[q]=0$ to be equivalent to $E(\lambda)[\varphi(q)]=0$  two things will have to happen.   First,  $\chi\cdot d\eta$ in \eqref{var} has to be an exact form. Second, it must be true that $E(\lambda)[\varphi(q)]\cdot\delta\varphi(q) = 0,\ \forall\ \delta q$ is  equivalent to the reduced equations $E(\lambda)[\varphi(q)] = 0$.  

With an additional technical hypothesis on the class of Lagrangians being considered, it is possible to give a very succinct set of necessary and sufficient conditions on the group action such that these two requirements are satisfied and the (the local version of) PSC holds. To state this result we need some definitions. Let ${\cal H}^{*}(G,G_{x})$ denote the Lie algebra cohomology of the the group $G$ relative to its isotropy subgroup $G_{x}$.  Let $p\in \kappa(E)$.  Let ${\rm Vert}_{p}(E)$ denote the vector space of vertical vectors in $T_{p}E$.  $G_{x}$ acts on ${\rm Vert}_{p}(E)$, where $x=\pi(p)$. Let $V_{p}\subset {\rm Vert}_{p}(E)$ denote the subspace of $G_{x}$-invariant vertical vectors. Let $V_{p}^{*} \subset {\rm Vert}^{*}_{p}(E)$ denote the vector space of $G_{x}$ invariant dual vectors. Let $V_{p}^{0} \subset {\rm Vert}^{*}_{p}(E)$ denote the annihilator of $V_{p}$. 

\bigskip\noindent
{\bf Theorem}

{\it The local version of PSC is valid for all $G$-invariant Lagrangians admitting a $G$-invariant boundary form if and only if, for each $x\in M$,

\medskip (i) ${\cal H}^{l}(G, G_{x}) \neq 0$,\quad $l$ is the dimension of the group orbits in $M$,

\medskip (ii) $V_{p}^{*} \cap V_{p}^{0} = 0$.}

\bigskip

Condition (i) only depends upon the abstract Lie group and its action upon the manifold $M$. Consequently it is independent of the nature of the fields being considered. Condition (ii), which is a local, pointwise version of the Palais condition, depends upon the group action on $E$, so depends upon the choice of fields. It is important to notice that this theorem characterizes PSC in terms of purely local features of the group action on $E$. Conditions (i) and (ii) are easily checked for a given group action.  If condition (i) fails to hold, there will exist Lagrangians for which some of the Euler-Lagrange equations of the reduced Lagrangian will not be correct. If condition (ii) fails to hold, there will exist Lagrangians for which the Euler Lagrange equations of the reduced Lagrangian  will fail to enforce all the reduced equations. 

The technical assumption we have introduced in this theorem is that the boundary form $\eta[\varphi, \delta\varphi]$ can be chosen to be $G$-invariant. This is very often the case in applications, {\it e.g.,} if one has a $G$-invariant metric on $M$ or a $G$-invariant connection on $E$, but it is not guaranteed in general. It is an interesting project to explore further the situation where no $G$-invariant $\eta$ exists.

Some important sufficient conditions for this version of PSC are worth mentioning.  First of all, as with Palais' PSC, both conditions (i) and (ii)  are satisfied if $G$ is a compact group.  Thus, {\it e.g.,} reduction of a Lagrangian by spherical symmetry will {\it always} work. Next, in the case where the group acts freely on $M$, condition (ii) is trivially satisfied and condition (i) reduces to the requirement that the symmetry group be unimodular.  Note that the  connected three dimensional Lie groups with Lie algebras of Bianchi class A are unimodular groups, so this special case explains the observation of Hawking.  More generally, if the group action is transverse on $E$, then again condition (ii) is trivially satisfied and the validity of PSC is controlled by the relative Lie algebra cohomology, in agreement with the results of Anderson and Fels \citep{AF:1997}.

I will finish this overview by providing a couple of very simple, brief illustrations of our theorem on PSC.  Given the ubiquitous role of general relativity in the history of this subject, it is only fitting that my examples involve this theory of gravity.  All the results shown in what follows were obtained using the {\sl DifferentialGeometry} package in {\sl Maple}. 

\section{Example: Homogeneous Spacetime}

We let $E$ be the bundle of Lorentz-signature metrics over a 4-dimensional manifold $M$.  Coordinates on $M$ are $x^{\mu}$, $\mu = 1,2,3,4$.  The group action is defined by a 5-dimensional subgroup $G$ of the diffeomorphism group of $M$ (acting on $E$ in the usual way) generated by the following vector fields on $M$: ($s\in{\bf R}$ is a parameter)
\begin{equation}
\label{vf1}
X_{1} = \frac{\partial}{\partial x^{2}},\quad  X_{2} = \frac{\partial}{\partial x^{3}},\quad X_{3} = - \frac{\partial}{\partial x^{1}} + x^{3} \frac{\partial}{\partial x^{2}}, \nonumber
\end{equation}
\begin{equation}
X_{4} = - x^{1} \frac{\partial}{\partial x^{1}} + x^{3} \frac{\partial}{\partial x^{3}}, \quad X_{5} = s x^{1}\frac{\partial}{\partial x^{1}} + s x^{2} \frac{\partial}{\partial x^{2}} + \frac{\partial}{\partial x^{4}}.
\end{equation}
This group is transitive on $M$.
  The isotropy subgroup at any point of $M$ is 1-dimensional, corresponding to a boost subgroup of the Lorentz group acting on the tangent space at the given point. This implies that obstruction (ii) to PSC is absent \citep{FT:2002}, as can also be checked explicitly.  
  
  We now check obstruction (i) to PSC, which involves computing relative Lie algebra cohomology.  In terms of a basis of left-invariant 1-forms $(\theta^{1}, \dots, \theta^{5})$ on $G$,  the structure equations for the Lie algebra of $G$ are
\begin{equation}
d\theta^{1} =  s \theta^{5} \wedge \theta^{1} - \theta^{2} \wedge \theta^{3},\ 
d\theta^{2} = \theta^{4}\wedge \theta^{2}, \ d\theta^{3} = \theta^{3}\wedge\theta^{4} - s \theta^{3}\wedge\theta^{5}, \ d\theta^{4}
 = 0, \ d\theta^{5} = 0.
 \end{equation}
Since $M$ is a homogeneous space with respect to $G$, the relative Lie algebra cohomology is the same at each point; we will perform our computations at the origin $x^{\alpha} = (0,0,0,0)$. The isotropy subgroup $H\subset G$ of  the origin is generated by the vector field $X_{4}$.  Since the group orbits are four-dimensional, we want to compute ${\cal H}^{4}(G,H)$. This cohomology can be obtained as follows. Let $\Omega^{*}$ be the set of all left-invariant forms $\omega$ on $G$ which are invariant under the right action of $H$ on $G$  and which satisfy $e_{4}\cdot \omega=0$, where $e_4$ generates the right action of $H$ on $G$. ${\cal H}^{4}(G,H)$ is the set of closed modulo exact 4-forms, all forms being taken from $\Omega^{*}$.    

Explicit computation reveals the following results. 
All elements of $\Omega^{4}$ are multiples of $\theta^{1}\wedge\theta^{2}\wedge\theta^{3}\wedge\theta^{5}$, which is a closed form.
All elements of $\Omega^{3}$ are linear combinations of $\theta^{1} \wedge \theta^{2} \wedge \theta^{3}$ and $\theta^{2} \wedge \theta^{3} \wedge \theta^{5}$, the latter being a closed form. We have
 \begin{equation}
d( \theta^{1} \wedge \theta^{2} \wedge \theta^{3})= - 2s\, \theta^{1}\wedge\theta^{2}\wedge\theta^{3}\wedge\theta^{5} .
\end{equation}
Thus, when $s\neq0$, PSC fails because of condition (i) in our theorem. PSC is satisfied when $s=0$. 

We can demonstrate this failure of PSC with the Einstein-Hilbert Lagrangian for a metric $g$,
\begin{equation}
\label{EH}
\lambda[g] = R \epsilon,
\end{equation}
where $R$ is the scalar curvature of the metric and $\epsilon$ is the volume form determined by the metric.  Using the covariant form of the metric as the field variable, the Euler-Lagrange expression is 
\begin{equation}
E(\lambda)[g] = {\cal E} \otimes \epsilon,
\end{equation}
where $\cal E$ is the (contravariant) Einstein tensor of $g$.
The most general $G$-invariant metric is given by 
\begin{equation}
\label{invg}
g = \frac{1}{2} d e^{-sx^{4}} dx^{1} \odot dx^{3} + c\omega \otimes \omega + \frac{b}{2} \omega \odot dx^{4} + a \,dx^{4}\otimes dx^{4},
\end{equation}
where $d>0$, $4ca - b^{2}>0$, and
\begin{equation}
\omega = e^{-sx^{4}}(dx^{2} + x^{1} dx^{3}).
\end{equation}
Thus the reduced ``fields'' are the parameters $(a, b, c, d)$.
All $G$-invariant skew tensors of type $\left(4\atop0\right)$ on $M$ are of the form
\begin{equation}
\chi=2 k e^{2sx^{4}} X_{1} \wedge X_{2} \wedge X_{3} \wedge X_{5},
\end{equation}
where $k\neq0$ is a constant.
The reduced Lagrangian for $(a, b, c, d)$ is then the 0-form:
\begin{equation}
\hat\lambda  =\frac{12kc (11d^2s^2-4ca+b^2)}{d\sqrt{4ca-b^2}},
\end{equation}
and its Euler-Lagrange expressions are obtained by simply differentiating with respect to $a$, $b$, $c$, $d$. 
There are 4 independent reduced equations, which can taken to be given by equating to zero the coordinate basis components (${\cal E}^{44}$, ${\cal E}^{24}$, ${\cal E}^{22}$, ${\cal E}^{13}$) evaluated on the metric \eqref{invg}.  We omit the expressions of these components, which are  a little lengthy. If PSC is valid, the vanishing of these four reduced equations should be equivalent to the vanishing of the derivatives of $\hat\lambda$ with respect to $(a, b, c, d)$, respectively.   Explicit computations reveal
\begin{eqnarray}
\frac{\partial\hat\lambda}{\partial a} - 24ke^{2sx^{4}}\sqrt{|g|}{\cal E}^{44} &=& -{384k c^{2} d \over(4ca-b^{2})^{3/2}} \, s^{2},\\
\frac{\partial\hat\lambda}{\partial b} - 24ke^{sx^{4}}\sqrt{|g|}{\cal E}^{24} &= & \frac{192 k  bcd}{(4ac - b^{2})^{3/2}} s^{2}\\
\frac{\partial\hat\lambda}{\partial c} - 24k\sqrt{|g|}{\cal E}^{22} &= &\frac{48kd (4ac-3b^{2})}{(4ac - b^{2})^{3/2}}s^{2}\\
\frac{\partial\hat\lambda}{\partial d} - 24k e^{sx^{4}}\sqrt{|g|}{\cal E}^{13} &= &  \frac{48kc }{(4ac - b^{2})^{1/2}}s^{2}.
\nonumber
\end{eqnarray}
As you can see, the reduced Einstein equations only agree with the Euler-Lagrange equations of the reduced Lagrangian when the symmetry group has $s=0$. This corresponds to the fact that the Lie algebra cohomology satisfies ${\cal H}^{4}(G,G_{x})\neq0$ only when $s=0$.

\section{Example: Plane Waves}

This example shows how condition (ii) in our Theorem comes into play (with a vengeance!).  We again consider the bundle of Lorentz-signature metrics over a 4-dimensional manifold $M$. Coordinates on $M$ are $(u,v,x,y)$. The group action is the canonical lift to $E$ of a 5-dimensional group of diffeomorphisms  of $M$ generated by the vector fields
\begin{equation}
X_{1} = \frac{\partial}{\partial v}, \ X_{2} = \frac{\partial}{\partial x}, \ X_{3} = \frac{\partial}{\partial y},\ 
X_{4}  = x \frac{\partial}{\partial v} + P(u) \frac{\partial}{\partial x},\ X_{5}  = y \frac{\partial}{\partial v} + Q(u) \frac{\partial}{\partial y},
\end{equation}
where $P(u)$ and $Q(u)$ are any functions such that $P^\prime(u)>0$, $Q^\prime(u)>0$.  The group action $\mu\colon G\times M\to M$ generated by these vector fields has 3-dimensional orbits $u=const.$; the isotropy subgroup of any point is two dimensional. While the group action  generated by these vector fields certainly depends upon the choice of the functions $P$ and $Q$, the abstract group $G$ does not. In terms of left-invariant forms $(\theta^{1},\dots,\theta^{5})$ the structure equations for the Lie algebra of $G$ are given by
\begin{equation}
d\theta^{1} = - \theta^{2}\wedge\theta^{4} - \theta^{3}\wedge\theta^{5},\ 
d\theta^{2}=0, \ 
d\theta^{3}=0, \ 
d\theta^{4}=0, \ 
d\theta^{5}=0. 
\end{equation}
The isotropy subgroup $G_{0}$ of a generic point $(u_{0}, v_{0},  x_{0}, y_{0})$ is generated by the Lie algebra spanned by the two vector fields $(V,W)$, where
\begin{equation}
V = X_{5} - y_{0} X_{1} - Q(u_{0}) X_{3},\quad W =  x_{0} X_{5} - y_{0} X_{4} + y_{0} P(u_{0}) X_{2} - x_{0} Q(u_{0}) X_{3}.
\end{equation}
A straightforward computation shows  that ${\cal H}^{3}(G, G_{0})\neq 0$, so condition (i) for PSC is satisfied.  

Now we examine condition (ii) for PSC for $G$ invariant Lagrangians built from a metric.  $V_{p}$ is the vector space of $G_{0}$ invariant rank-2, symmetric tensors at $(u_{0}, v_{0},  x_{0}, y_{0})\in M$; it is spanned by two quadratic forms:
\begin{equation}
{\cal Q}_{1} = du \otimes du, \quad {\cal Q}_{2} = -Q^{\prime}(u_{0}) P^{\prime}(u_{0}) du \odot dv + Q^{\prime}(u_{0}) dx \otimes dx + P^{\prime}(u_{0}) dy \otimes dy.
\end{equation}
Similarly, the vector space $V^{*}_{p}$ is spanned by
\begin{equation}
\frac{\partial}{\partial v} \otimes \frac{\partial}{\partial v}, \quad \frac{\partial}{\partial u} \odot \frac{\partial}{\partial v} - P^{\prime}(u_{0}) \frac{\partial}{\partial x} \otimes \frac{\partial}{\partial x} - Q^{\prime}(u_{0}) \frac{\partial}{\partial y} \otimes \frac{\partial}{\partial y}.
\end{equation}
$V_{p}^{0}$ is the vector space of symmetric tensors of type $\left(2\atop0\right)$ at $(u_{0}, v_{0},  x_{0}, y_{0})$ which map ${\cal Q}_{1}$ and ${\cal Q}_{2}$ to zero upon contraction of all arguments. It is clear that $\frac{\partial}{\partial v} \otimes \frac{\partial}{\partial v}\in V_{p}^{0}$ and is thus a non-zero element of $V_{p}^{*} \cap V_{p}^{0}$. In fact this tensor spans the intersection. Thus condition (ii) for PSC is violated. 

We can see how PSC fails via the Einstein-Hilbert Lagrangian \eqref{EH}. The most general $G$-invariant metric takes the form
\begin{equation}
\label{ginv}
g = a(u) {\cal Q}_{1} + b(u) {\cal Q}_{2},
\end{equation}
where $b(u)>0$.  The reduced fields are $a(u)$ and $b(u)$.  
The Einstein tensor is of the form
\begin{equation}
{\cal E} = \Delta[b] D_{v}\otimes D_{v},
\end{equation}
where $\Delta[b]$ is a non-linear second order differential operator, built from $P$ and $Q$, acting upon $b(u)$.  I will not show $\Delta[b]$ explicitly because it is a bit of a mess and, in any case, is not needed for this discussion.  Thus the reduced field equations leave $a(u)$ arbitrary and impose the condition $\Delta[b] = 0$ on $b(u)$. The solution of this equation determines gravitational plane waves \citep{Bondi:1959, CGT:2006}.
If we evaluate the Einstein-Hilbert Lagrangian on $g$ in \eqref{ginv} we find it vanishes identically --- this means the reduced Lagrangian is zero!\footnote{More generally, it can be shown that any diffeomorphism invariant Lagrangian will yield a trivial Lagrangian when evaluated on a metric of the form \eqref{ginv} \citep{CGT:2006}.}  Thus  the reduced Lagrangian fails to yield the reduced equation $\Delta[b]=0$.
This is the way in which PSC fails when condition (ii) is not satisfied --- one or more of the reduced equations will fail to  appear as Euler-Lagrange equations of the reduced Lagrangian.  

\bigskip\noindent
{\bf Acknowledgement}

I would like to thank the organizers of the XIX International Fall Workshop on Geometry and Physics for granting me the privilege of participating.

\bibliographystyle{aipproc}
\bibliography{PSC_Porto}

\begin{thebibliography}{15}
\expandafter\ifx\csname natexlab\endcsname\relax\def\natexlab#1{#1}\fi
\providecommand{\enquote}[1]{``#1''}
\expandafter\ifx\csname url\endcsname\relax
  \def\url#1{\texttt{#1}}\fi
\expandafter\ifx\csname urlprefix\endcsname\relax\def\urlprefix{URL }\fi
\providecommand{\eprint}[2][]{\url{#2}}

\bibitem[Palais(1979)]{Palais:1979}
R.~Palais, \emph{Commun. Math. Phys.} \textbf{69}, 19--30 (1979).

\bibitem[Olver(1993)]{Olver:1993}
P.~Olver, \emph{Applications of Lie Groups to Differential Equations},
  Springer-Verlag, 1993.

\bibitem[Anderson et~al.(2000)]{AFT:2000}
I.~M. Anderson, M.~E. Fels, and C.~G. Torre, \emph{Commun. Math. Phys.}
  \textbf{212}, 653--686 (2000), \eprint{math-ph/9910015}.

\bibitem[Barbero and Villase\~nor(2010)]{lrr-2010-6}
F.~Barbero, and E.~Villase\~nor, \emph{Living Rev. Relativity} \textbf{13}, 6
  (2010).

\bibitem[Anderson and Fels(1997)]{AF:1997}
I.~M. Anderson, and M.~E. Fels, \emph{American Journal of Mathematics}
  \textbf{119}, 609 (1997).

\bibitem[Anderson et~al.(1999)]{Anderson:1999cm}
I.~M. Anderson, M.~E. Fels, and C.~G. Torre, \emph{CRM Proceedings \& Lecture
  Notes} \textbf{29}, 95--108 (1999), \eprint{math-ph/9910014}.

\bibitem[Fels and Torre(2002)]{FT:2002}
M.~E. Fels, and C.~G. Torre, \emph{Class. Quant. Grav.} \textbf{19}, 641--676
  (2002), \eprint{gr-qc/0108033}.

\bibitem[Weyl(1917)]{Weyl:1917}
H.~Weyl, \emph{Ann. Phys., Lpz.} \textbf{54}, 117 (1917).

\bibitem[Lovelock(1973)]{Lovelock:1973}
D.~Lovelock, \emph{Nuovo Comento B} \textbf{73}, 260 (1973).

\bibitem[Hawking(1969)]{Hawking:1969}
S.~Hawking, \emph{Mon. Not. R. Astron. Soc.} \textbf{142}, 129 (1969).

\bibitem[MacCallum and Taub(1972)]{MacCallum:1972}
M.~MacCallum, and A.~Taub, \emph{Commun. Math. Phys.} \textbf{25}, 173 (1972).

\bibitem[Ryan(1974)]{Ryan:1974}
M.~Ryan, \emph{J. Math. Phys.} \textbf{15}, 812 (1974).

\bibitem[Sneddon(1976)]{Sneddon:1976}
G.~Sneddon, \emph{J. Phys. A: Math. Gen.} \textbf{9}, 229 (1976).

\bibitem[Bondi et~al.(1959)]{Bondi:1959}
H.~Bondi, F.~Pirani, and I.~Robinson, \emph{Proc. Roy. Soc. London A}
  \textbf{251}, 519 (1959).

\bibitem[Torre(2006)]{CGT:2006}
C.~G. Torre, \emph{Gen. Rel. Grav.} \textbf{38}, 653--662 (2006).

\end{thebibliography}

\end{document}